\documentclass[11pt,preprint]{aastex}
\usepackage{graphicx}
\newcommand{\lsim}{\mathrel{\hbox{\rlap{\lower.55ex\hbox{$\sim$}} 
\kern-.3em \raise.4ex \hbox{$<$}}}}
\newcommand{\gsim}{\mathrel{\hbox{\rlap{\lower.55ex\hbox{$\sim$}} 
\kern-.3em \raise.4ex \hbox{$>$}}}}

\slugcomment{Accepted by the Astrophysical Journal, 2005 Feb. 15}

\shortauthors{Boboltz \& Diamond}
\shorttitle{SiO Masers Toward IK\,Tau}

\begin{document}

\title{Axial Symmetry and Rotation in the SiO Maser Shell of IK\,Tauri} 
\author{D. A. Boboltz}
\affil{U.S. Naval Observatory,\\
3450 Massachusetts Ave., NW, Washington, DC 20392-5420 \\
dboboltz@usno.navy.mil}
\and
\author{P. J. Diamond}
\affil{Jodrell Bank Observatory, University of Manchester,\\
Macclesfield, Cheshire SK11 9DL, UK \\
pdiamond@jb.man.ac.uk}

\begin{abstract}  
We observed $v=1, J=1-0$ 43-GHz SiO maser emission toward the 
Mira variable IK\,Tauri (IK\,Tau) using the Very Long Baseline Array (VLBA).
The images resulting from these observations show that SiO masers
form a highly elliptical ring of emission approximately $58 \times 32$\,mas
with an axial ratio of 1.8:1.  The major axis of this elliptical
distribution is oriented at position angle of $\sim$59$^{\circ}$.
The line-of-sight velocity structure of the SiO masers has an  
apparent axis of symmetry consistent with the elongation axis of
the maser distribution.  Relative to the assumed stellar velocity of 
35\,km\,s$^{-1}$, the blue- and red-shifted masers were found to lie to the 
northwest and southeast of this symmetry axis respectively.  This 
velocity structure suggests a NW--SE rotation of the SiO maser 
shell with an equatorial velocity, which we determine to be 
$\sim$3.6\,km\,s$^{-1}$.   Such a NW--SE rotation is in agreement 
with a circumstellar envelope geometry invoked to explain 
previous H$_2$O and OH maser observations.  In this geometry,
H$_2$O and OH masers are preferentially created in a region of 
enhanced density along the NE--SW equator orthogonal 
to the rotation/polar axis suggested by the SiO maser velocities.

\end{abstract}


\section{INTRODUCTION}

The star IK\,Tau, also know as NML Tau \citep{NML:65}, 
is an extremely red Mira variable with an unusually late 
spectral type ranging from M8.1--M11.2 \citep{WL:73}.  The period of
IK\,Tau is $\sim$470 days as determined from infrared photometry 
measurements at 1\,$\mu$m
\citep{WL:73} and at 11\,$\mu$m \citep{HBDHLMTTJLG:97}.  
\cite{OLNW:98} computed a distance of 250\,pc from integrated visual,
near-infrared and IRAS data using the period-luminosity relationship
of \cite{WMFMCRCC:94}.  This distance is consistent with the 270\,pc 
determined by \cite{HBFN:72} using infrared flux measurements and
an assumed long-period variable maximum luminosity of 
$10^{-4}$.  Mass loss rates estimated from the CO($J=1-0$)
line range from $5.1 \times 10^{-6} M_{\odot}$\,yr$^{-1}$ \citep{KM:85}
to $2.4 \times 10^{-6} M_{\odot}$\,yr$^{-1}$ \citep{OLNW:98}.
  
Long-period variables such as IK\,Tau often exhibit circumstellar maser
emission in one or more molecular species (i.e. OH, H$_2$O or SiO).
IK\,Tau is one such star that has maser emission in all three species.
The OH and H$_2$O masers toward IK\,Tau have been mapped using
connected element and radio-linked interferometers
\citep{LJBSD:87,BJD:89,BCJ:93,YC:94,BCLRRY:03} revealing the 
structure of the
circumstellar envelope (CSE) on both large scales ($2''-4''$ traced by
the OH masers) and intermediate scales (200--350 mas traced by the
H$_2$O masers).  Such studies have been useful in determining the 
structure of the CSE on these scales.  The SiO masers, which lie close 
to the star, provide a unique probe of the morphology of the CSE and 
the velocity of the gas in a region just a few stellar radii from the
photosphere.  Previous Very Long Baseline Interferometry (VLBI)
observations of long-period variables have demonstrated the various 
morphologies that exist in the SiO maser region.   Some stars exhibit 
ring-like SiO maser distributions \citep{DKJZBD:94,DK:98, COTTON:04},
while others (e.g. NML\,Cyg) show more elliptical distributions \citep{BM:00}.
Previous SiO maser studies have also
revealed the velocity structure of the nearby CSE to be complex showing
elements of contraction \citep{BDK:97}, expansion \citep{DK:03}, 
and even rotation \citep{BM:00, HBPWF:01, SC:02, COTTON:04}.  
In this article, we present the results from our VLBI observations of the SiO maser 
emission toward the long-period variable IK\,Tau.  We discuss how 
these observations fit into the overall picture of the CSE as traced by 
the various maser species and the possibility of axial symmetry and 
rotation in the envelope of IK\,Tau.

\section{OBSERVATIONS AND REDUCTION}

We observed the $v=1, J=1-0$, SiO maser emission toward IK\,Tau 
($\alpha = 03^h 53^m 28^{s}.8$, $\delta = 11^{\circ} 24' 22''.4$, J2000) 
on 1996 April 5 from 20:00 to 24:00 UT.  Based on a period of 470
days \citep{WL:73, HBDHLMTTJLG:97} and a maximum at JD\,2439440
\citep{WL:73} the stellar phase of IK\,Tau was $\sim$0.85 during the
epoch of our observations.  IK\,Tau and a continuum calibrator
(0423-013) were observed using the 10 stations of the VLBA plus a
single antenna of the Very Large Array (VLA).  Both the VLBA and VLA
are operated by the National Radio Astronomy Observatory
(NRAO).\footnote{The National Radio Astronomy Observatory is a
facility of the National Science Foundation operated under cooperative
agreement by Associated Universities, Inc.}  A reference frequency 
of 43.122027\,GHz was used for the $v=1, J=1-0$ SiO transition.  
Data were recorded in dual circular polarization using a single 
4-MHz (27.8\,km\,s$^{-1}$) band centered on the local standard 
of rest (LSR) velocity of 34.0\,km\,s$^{-1}$.  System temperatures 
and point source sensitivities were on the order of $\sim$150\,K 
and $\sim$11\,Jy\,K$^{-1}$ respectively.

The data were correlated at the VLBA correlator operated by the NRAO in 
Socorro, New Mexico.  Auto and cross-correlation
spectra consisting of 128 channels with channel spacings of 31.25\,kHz 
($\sim$0.2\,km\,s$^{-1}$) were produced by the correlator.  Calibration
was performed using the Astronomical Image Processing
System (AIPS) maintained by NRAO.  Total intensity (Stokes $I$) data
were calibrated in accordance with the procedures outlined in
\citet{DIAMOND:89}.  The bandpass response was determined from scans
on the continuum calibrator and used to correct the target source
data. The time-dependent gains of all antennas relative to a reference
antenna were determined by fitting a total-power spectrum (from the
reference antenna with the target source at a high elevation) to the
total power spectrum of each antenna.  The absolute flux density scale
was established by scaling these gains by the system temperature and
gain of the reference antenna.  Errors in the gain and pointing of the
reference antenna and the atmospheric opacity contribute to the error
in the absolute amplitude calibration, which is accurate to about
15--20\%.  Residual group delays were determined by performing 
a fringe fit to the continuum calibrator sources and were applied 
to the spectral line data.   Residual fringe-rate solutions were obtained 
by fringe-fitting a strong reference feature in the channel at 
$V_{\rm LSR} = 36.2$\,km\,s$^{-1}$ and were applied to all channels 
in the spectrum.  An iterative self-calibration and imaging procedure 
was performed to map the reference channel at 36.2\,km\,s$^{-1}$.  
The resulting phase solutions were applied to all channels in the band.

\begin{figure}[hbt]
\includegraphics[width=4in]{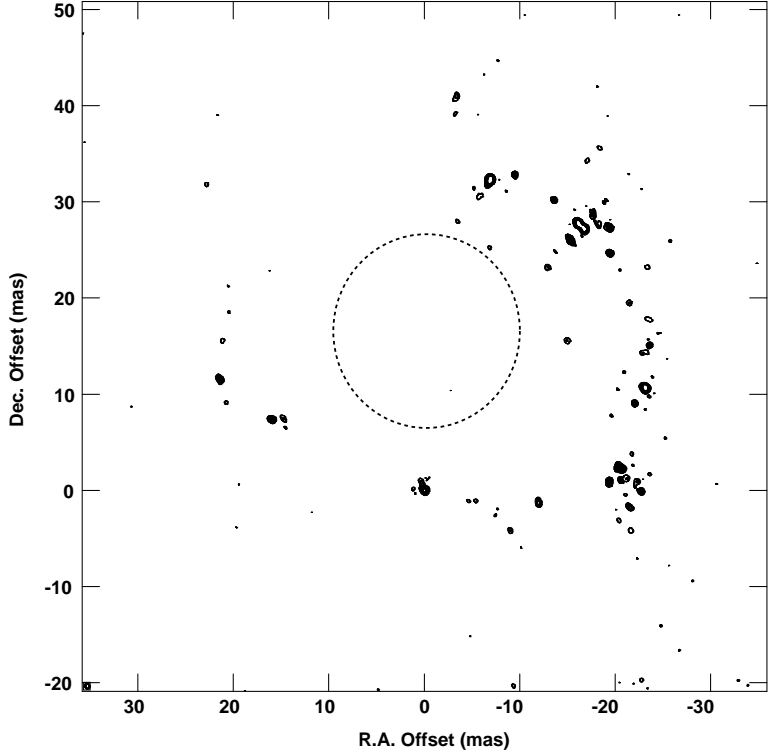}
\figcaption{Total intensity VLBI images of the $v=1, J=1-0$ SiO maser 
emission toward IK\,Tau integrated over the LSR velocity range from 
$+28.8$\,km\,s$^{-1}$ to $+41.6$\,km\,s$^{-1}$.  Contour levels are 
-3, -1.5, 1.5, 3, 5, 10, 20, 40, and 80\% of the peak integrated 
flux density of 49.9~Jy~beam$^{-1}$.  The synthesized beam is 
$0.54\times 0.41$\,mas at a position angle of 14.1$^{\circ}$.  The 
dashed circle in the represents a stellar uniform disk diameter of 
20.2\,mas from \citep{MONNIER:04}.
The location of the stellar disk is conjecture. \label{IKTAU_ISUM}}
\end{figure}

Images of the SiO maser emission consisting of $1024\times 1024$ pixels 
($\sim$$72\times72$\,mas) were generated using a synthesized beam of 
$0.54\times 0.41$\,mas in natural weighting.  Images were produced for 
spectral channels from 28.8\,km\,s$^{-1}$ to 41.6\,km\,s$^{-1}$ forming an
image cube of 60 planes.  Off-source RMS noise estimates in the spectral 
channel images range from 10\,mJy\,beam$^{-1}$ in planes with weak or no 
maser emission to 80\,mJy\,beam$^{-1}$ in dynamic-range limited planes 
containing strong ($\gsim$15\,Jy\,beam$^{-1}$) maser emission.   
Figure \ref{IKTAU_ISUM} shows the total intensity 
contour map of the 43.1-GHz SiO masers integrated over the LSR velocity 
range above.  The dashed circle in the figure represents the stellar 
photosphere with a diameter of 20.2\,mas.  This diameter was determined
from a uniform disk (UD) diameter fit to combined Keck 
aperture-masking data and Infrared-Optical Telescope Array (IOTA) data 
in the near infrared \cite{MONNIER:04}.  It is apparent from 
Figure \ref{IKTAU_ISUM} that the 
SiO masers toward IK\,Tau form an elongated ring roughly $58\times32$\,mas 
across oriented in the NE--SW direction.   If the star is assumed to be in the 
center of the distribution, maser distances range from $\sim$1.6--2.8 stellar 
radii.

In order to identify and parameterize maser components, two-dimensional
Gaussian functions were fit to the emission in each spectral (velocity) 
channel using the AIPS task SAD.   Image quality was assessed using
the off-source RMS noise and the deepest negative pixel in the image.
A cutoff, above which emission features were fit, was conservatively 
set to the greater of 10$\sigma_{\rm RMS}$ or 80\% of the absolute 
value of the deepest negative pixel in the plane.  Errors reported 
by SAD were based on the source size divided by twice the signal-to-noise 
ratio (SNR) in the image and 
ranged from 5\,$\mu$as for features with high SNR, to 371\,$\mu$as
for features with lower SNR.  The emission features identified by
SAD are represented by the circles in Figure \ref{IKTAU_SPOT}.  In the figure, 
point sizes are are proportional to the logarithm of the fitted flux 
density which ranged from 0.2 to 18.3\,Jy.  Comparing Figures 1 and 2, 
we see that the identified features accurately represent the emission 
summed over all velocity channels in the image cube.  

\section{RESULTS AND DISCUSSION}

\subsection{Axial Symmetry in the CSE of IK\,Tau}

\begin{figure}[hbt]
\plotone{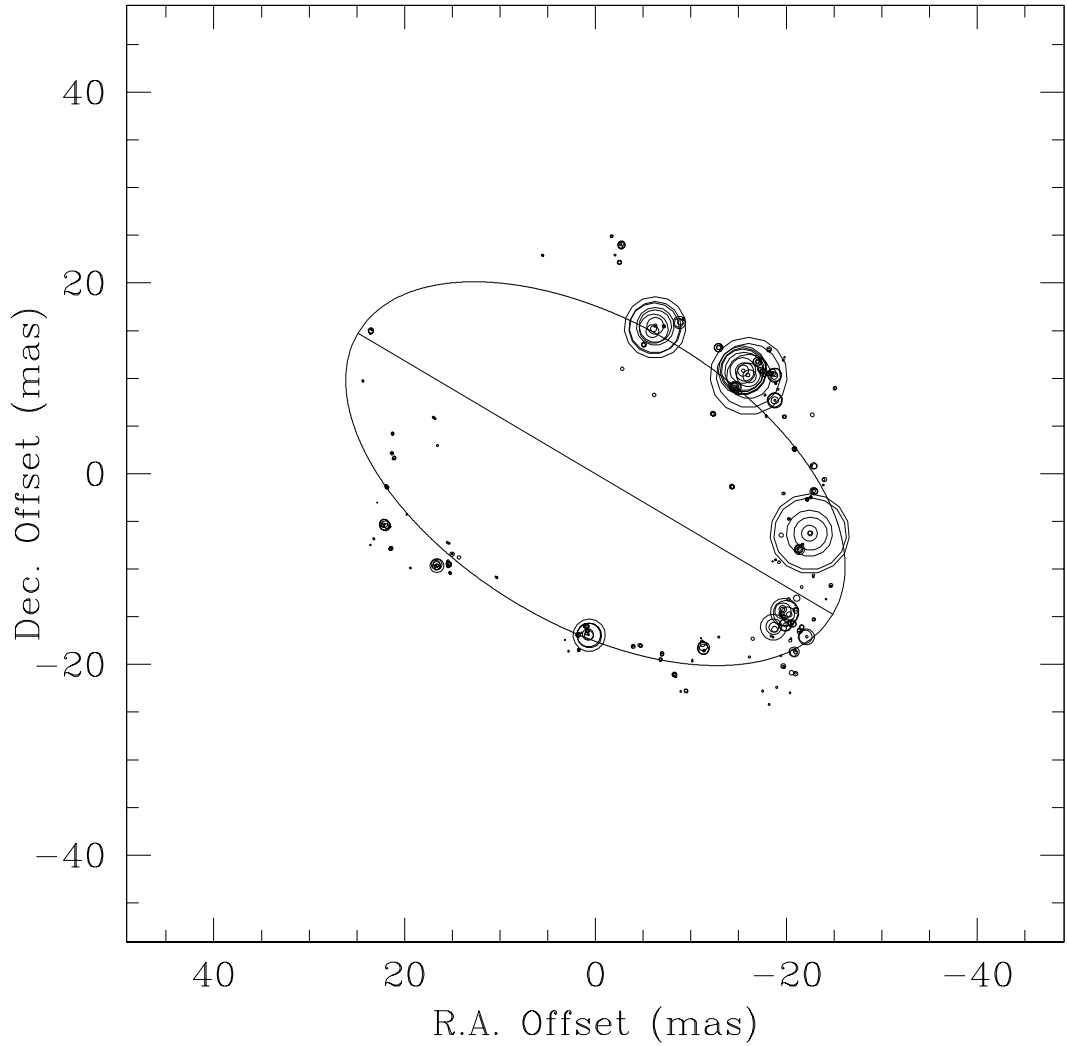}
\figcaption{Maser spot map as derived from the two-dimensional Gaussian
fits to masers in each plane in the image cube.  Point sizes are  
proportional to the logarithm of the flux density.  The ellipse indicates 
a flux-density weighted least-squares fit to the maser distribution.
\label{IKTAU_SPOT}}
\end{figure}

Figures \ref{IKTAU_ISUM} and \ref{IKTAU_SPOT} show that the 
distribution of the 43.1-GHz SiO maser
emission toward IK\,Tau was clearly not circularly symmetric during the 
epoch of our observations.  Instead, the distribution appears to be 
elongated along an axis running from the northeast to the southwest in
the images.   We characterized this morphology by performing a least
squares fit of an ellipse to the distribution of masers weighted by
the flux density of each component.  The resulting ellipse is plotted
in Figure \ref{IKTAU_SPOT}.  The lengths of the semimajor and semiminor 
axes were found to be 29\,mas and 16\,mas respectively with the semimajor axis
of the ellipse oriented at 59$^{\circ}$ measured east of north.  At
the assumed distance to IK\,Tau of 250\,pc, the distribution of SiO
masers is approximately $15\times 8$\,AU.   The SiO maser distributions
of $\lsim$20 late-type stars have previously been imaged with 
angular resolutions provided by VLBI \citep[e.g.][]{COTTON:04, DK:03, SC:02, 
HBPWF:01, BM:00, DKJZBD:94}.  Of these stars, 
IK Tau, with an axial ratio of $\sim$1.8:1, appears to have one of the most 
elongated SiO rings yet observed. 
  
As mentioned previously, \cite{MONNIER:04} were able to separate the 
near-infrared stellar and dust contributions and determine a
UD diameter of 20.2\,mas for the photosphere of IK\,Tau.  However, they 
found that the longest (27\,m) IOTA baseline data were inconsistent with 
this fit, and the data were thus ignored.  Possible explanations for this 
deviation are departures from uniform photospheric brightness (e.g. stellar 
hot spots or extended molecular emission), or the presence of a binary 
companion \citep{MONNIER:04}.    

The circumstellar dust shell, typically outside the SiO maser region,
has been observed for IK\,Tau using long baseline interferometers operating 
at mid-infrared \citep{DBDGT:94,HBDHLMTTJLG:97} and near-infrared
\citep{MONNIER:04} wavelengths.  
\cite{HBDHLMTTJLG:97} found that their Infrared Spatial 
Interferometer (ISI) 11\,$\mu$m visibilities were best fit by a three 
spherically symmetric concentric dust shells.  However, \cite{MONNIER:04} 
state that their Keck-IOTA data, which includes multiple late-type stars,
suggest that the popular strategy of incorporating multiple spherically 
symmetric shells to model deviations from simple uniform outflow 
might be misguided and that clumpiness and global asymmetry should 
be considered more seriously.   In fact, \cite{MONNIER:04} report that 
the visibility data for IK\,Tau shows signs of asymmetries for the dust shell 
but the data are not of sufficient quality to make a definitive determination.  

Beyond the regions of the SiO masers and circumstellar dust shell, 
there is evidence for axisymmetric structures in the CSE of IK\,Tau.  
On large angular scales,
\citet{BJD:89} found that the 1612-MHz OH masers form an elongated
ring $\sim$3.5$''\times2.3''$ with a NW--SE orientation of the long
axis.  The axisymmetry is especially apparent at high velocities
($V>45$\,km\,s$^{-1}$) where nearly all of the emission originates to
the SE and NW of the star.  Observations of the 22-GHz H$_2$O masers
with the VLA \citep{LJBSD:87,BCJ:93} and the 
Multi Element Radio Linked Interferometer Network (MERLIN) 
\citep{YC:94,BCLRRY:03} show similar axial symmetry 
with respect to the NW--SE direction.  VLA observations by \cite{LJBSD:87} 
show that the low-velocity masers have a ring-like distribution, while the  
high-velocity masers have a flattened distribution with a NW--SE 
orientation.  The overall size of the shell was roughly 330\,mas.
The VLA data of \cite{BCJ:93} show the H$_2$O maser distribution to
be $\sim$220\,mas across with a similar elongation in the NW--SE 
direction.   The MERLIN images of \cite{BCLRRY:03} show bright
blue- and red-shifted emission distributed over an elliptical
region approximately 100$\times$~200mas superimposed on a fainter
spherical region of emission from all spectral channels.

The observed NE--SW orientation for the elliptical SiO maser distribution
is roughly orthogonal to elongations of the H$_2$O and OH maser
distributions.   \citet{BJD:89} suggest that the axisymmetric structures seen 
in the OH and H$_2$O maser shells for certain late-type stars are a result of
the density distribution of gas and dust in the CSE.  They conclude
that the density distribution is latitude dependent with the total
density decreasing as the latitude increases from the rotationally 
defined equatorial plane.  In stars in which both the OH and 
H$_2$O masers are axially symmetric (e.g. IK\,Tau) \citet{BJD:89} 
suggest that the distributions
should be aligned parallel along the equatorial plane since the
density enhancement is conducive to the formation of the H$_2$O masers
(assuming these masers are collisionally pumped) and since the OH
masers are the result of the photodissociation of the H$_2$O by the
surrounding UV radiation.  \citet{YC:94} state that their MERLIN
data are also supportive of a NW--SE equator for IK\,Tau with
the brightest H$_2$O masers observed along this axis as expected.
Assuming this NW--SE axis is parallel to the equator of IK\,Tau as
suggested by the OH and H$_2$O maser observations, then the elongation 
we observe in the SiO maser distribution is along an axis roughly 
perpedicular to the equator and possibly parallel to the polar axis 
of the star.  The importance of this geometry will become apparent 
in the discussion of the velocity structure of the SiO masers in the 
next section.

\subsection{Velocity Structure of the SiO Masers}

\begin{figure}[hbt]
\plotone{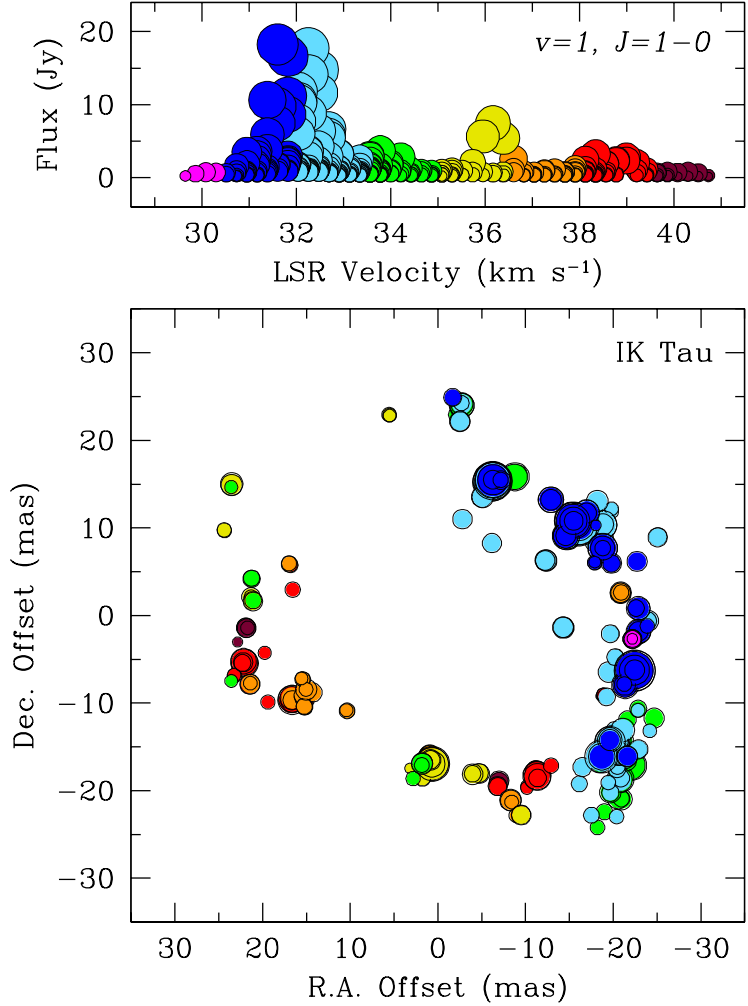}
\figcaption{LOS velocity structure of the SiO maser emission toward
IK\,Tau as measured by the two-dimensional Gaussian fit to the masers 
in the VLBI channel maps.  The top panel shows the spectrum formed 
by plotting component flux density versus velocity, color coded in 
1.5\,km\,s$^{-1}$ velocity increments from redward (left) to blueward
(right).  The bottom panel plots the spatial and velocity 
distribution of the masers.  The color of each point represents 
the corresponding velocity bin in the spectrum and the size of each 
point is proportional to logarithm the flux density.  
Errors in the positions of the features are smaller than the data 
points. \label{IKTAU_VEL}}
\end{figure}

The line-of-sight (LOS) velocity information derived from our VLBA
data is shown in Figure \ref{IKTAU_VEL}.  The top panel shows the spectrum of SiO
maser emission from 28--42\,km\,s$^{-1}$ color coded by LOS
velocity in increments of 1.5\,km\,s$^{-1}$.  The bottom panel of 
Figure \ref{IKTAU_VEL} shows the SiO masers plotted in the same 
velocity color coding as the top panel, with the color of the maser spot 
representing its corresponding velocity in the spectrum.  It is apparent from the
figure that there is a symmetry axis for the velocity structure of the
SiO maser emission toward IK\,Tau with the northwestern side of the
shell dominated by blue-shifted masers (relative to the assumed stellar
velocity of 35\,km\,s$^{-1}$) and the southeastern side of the shell
dominated by red-shifted masers.  Such velocity distributions in 
which blue and red-shifted SiO masers occur on opposite sides of the 
star have previously been observed for NML\,Cyg \citep{BM:00}, 
R\,Aqr \citep{HBPWF:01,COTTON:04}, OH231.8+4.2 \citep{SC:02}.  
For NML\,Cyg and R\,Aqr, the maser velocity distributions suggested
rotation of their respective maser shells.  In the case of OH231.8+4.2 
\cite{SC:02} concluded that the maser distribution could be modeled 
by a rotating, contracting torus of emission. 

The ring-like distributions typically observed in VLBA studies 
of SiO masers indicate the masers are amplified in a direction 
perpendicular to the radial direction (i.e. tangential amplification) 
\cite{DKJZBD:94}.  This amplification occurs along the longest path lengths
parallel to the observers line of sight.   Assuming this is 
the case for IK\,Tau, then the velocity structure shown in
Figure \ref{IKTAU_VEL} suggests a NW to SE rotation of the maser shell with the axis
of rotation near the elongation axis of the distribution.  Such a
rotation axis is in agreement with a NE--SW equator for IK\,Tau as
indicated by the previous OH and H$_2$O maser observations mentioned
above.

Rotation is not a new concept with regards to the masers in the
CSE of IK\,Tau.  \citet{BJD:89} considered models
with radial expansion combined with Keplerian rotation to explain the
axisymmetric geometries observed for the 1612 MHz OH masers.  They
concluded that although rotation could not unequivocally be ruled out,
the data did not support rotation as a primary component in the
kinematics of the OH shell.  Additionally, \citet{BCJ:93} found that
the plot of radius versus velocity for their H$_2$O maser data was
reminiscent of models of accelerating or axisymmetric outflow with a
possible rotational component.

To better characterize the velocity structure of the SiO emission, we
investigated a number of models with velocity components including
rotation and expansion/contraction.  Modeling the three-dimensional
velocity structure of the SiO masers toward IK\,Tau is complicated by
the elliptical distribution of the masers and by a lack of knowledge
of the shell structure along the observer's line of sight (LOS). 
The elliptical distribution of the masers could be modeled by either 
a ellipsoidal shell or an inclined torus similar to that of \cite{SC:02}.  
We chose an ellipsoidal shell based on the assumption that the masers 
are tangentially amplified and the fact that nearly a full ring of
emission is observed.  In the case of a torus with tangential amplification
one would expect to see only two regions of emisison on opposite sides 
of the star as was observed by \cite{SC:02}.  Models developed for IK\,Tau
are based on a prolate spheroid with the long axis oriented at a position angle of 
59$^\circ$ (as suggested by the elliptical fit shown in Figure \ref{IKTAU_SPOT}).  
Since the true three-dimensional shape of the maser shell 
is unknown, the axis of the ellipsoidal model along the observers LOS
was assumed to be the same size as the minor axis of the model in the
plane of the sky.  Additionally, the inclination of the ellipsoid was
assumed to be zero.  

\begin{figure}[hbt]
\centering 
\plotone{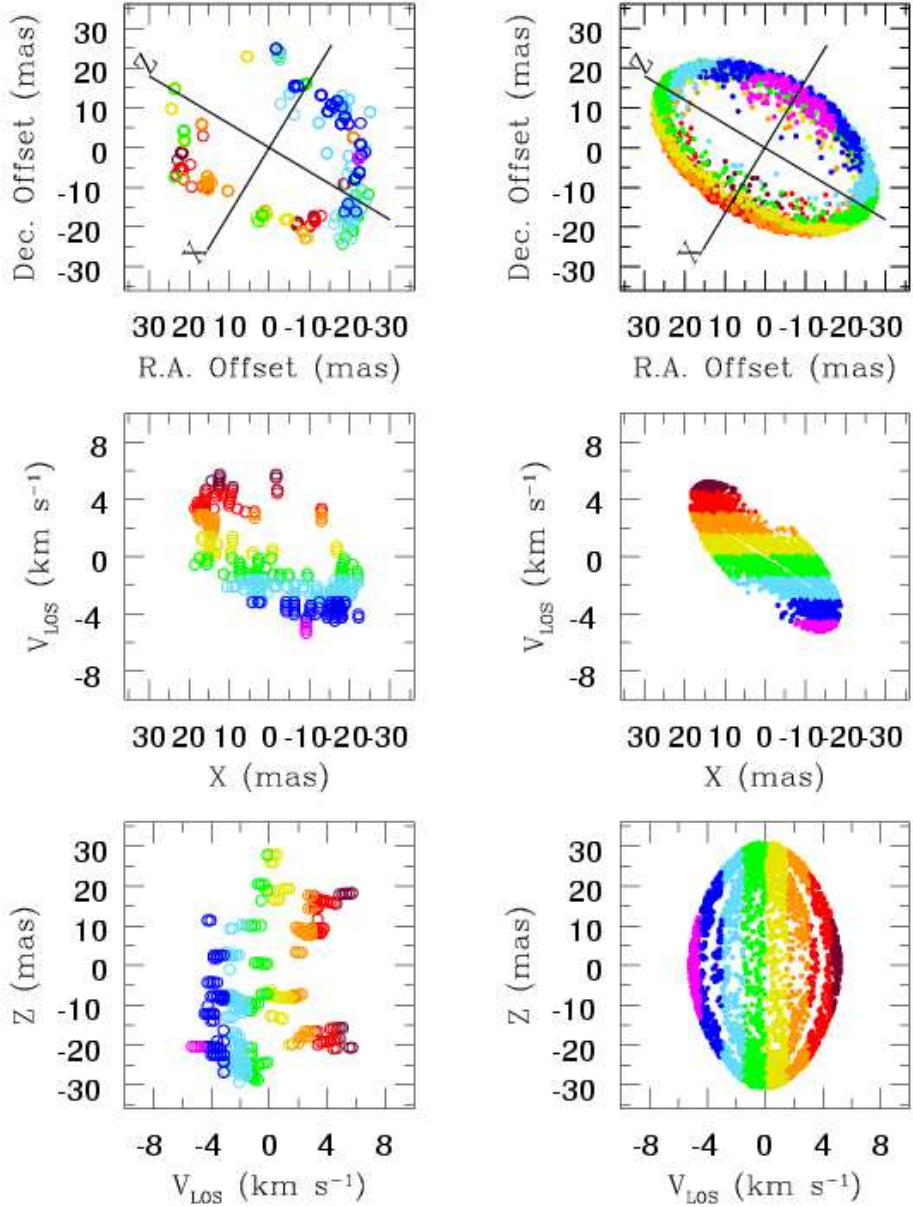}
\figcaption{Side-by-side color-coded plots of the VLBA data (left)
and the ellipsoidal shell model (right).  Velocity color coding is the same as for 
Figure \ref{IKTAU_VEL} except that the assumed LSR velocity of IK\,Tau 
($V_{\rm LPV} = 35$\,km\,s$^{-1}$) has been subtracted.  The top panels show
the orientation of the VLBA data (left) and model (right) 
distributions on the sky.  The top panel shows the assumed rotational
and equatorial axes, Z and X respectively, with the rotation axis
at P.A. = 59$^{\circ}$.  The top left panel is analogous to the 
Figure \ref{IKTAU_VEL} bottom panel.  The middle and bottom panels show the 
LOS velocities plotted against the X and Z axes of the 
distribution respectively for both the VLBA data (left) 
and the model (right). \label{IKTAU_MODEL}} 
\end{figure}

The models were compared ``by eye" to the actual data to determine how 
well they reproduced the true maser spatial and velocity distributions.   
To make the comparison easier we plotted the data and the models in 
a manner similar to that presented in \cite{HBPWF:01} with three different 
views.  Figure \ref{IKTAU_MODEL} shows the VLBA data plotted in the 
three panels on the left.  The top panel plots the maser distribution with 
the same color-coding as that of Figure \ref{IKTAU_VEL} along with a set 
of axes ($X$,$Z$) determined from the elliptical fit of Figure \ref{IKTAU_SPOT}.  
The middle panel 
plots the component LOS velocity versus its position along the
equatorial ($X$) axis.  The bottom panel plots the component LOS velocity 
versus its position along the polar ($Z$) axis.  The right three panels of
Figure \ref{IKTAU_MODEL} plot the equivalent distributions derived from the model.

The size and shape of the model shell (i.e. major axis, minor axis and shell 
thickness) were chosen to approximate the spatial distribution of the SiO 
masers as seen in Figure \ref{IKTAU_VEL} and the top-left panel of Figure \ref{IKTAU_MODEL}.  Maser velocities were first modeled with a general 
form of rotation given by:
\begin{equation}
V(r) \propto A r^p.
\end{equation}
It quickly became apparent that models including only rotation
do not sufficiently approximate the maser data.
In the middle-left panel of Figure \ref{IKTAU_MODEL}, we see that there is a 
significant variation in position along the $X$-axis for masers 
with small ($<$2\,km\,s$^{-1}$) LOS velocities.  It is this spread in
position for low-velocity points that is difficult to model assuming 
only rotation.   For example, solid-body rotation ($p=1$) yields 
a simple linear variation of LOS velocity with position along the 
$X$-axis regardless of distance along the $Z$-axis.  Keplerian 
rotation ($p=-0.5$) gives a greater variation in $X$-axis position 
for higher velocity masers, however, at lower velocities the positions 
are still confined to a narrow region in the $V_{\rm LOS}$ vs. $X$
plot (see Figure 3 of \cite{HBPWF:01}).  

One way to produce the observed spread in position at small LOS velocities
is to add an expansion term to the equation describing the motion of the 
masers in the shell.   The LOS velocity for an elliptical shell which is in rotation 
and also has a component of constant velocity expansion, $V_{\rm exp}$,
can be written:
\begin{equation}
V(r)_{\rm LOS} = V_{\rm LPV} +  \sqrt{GM \over r^{q}} \cos \theta \cos \phi 
+ V_{\rm exp} \sin \theta \cos \phi
\end{equation}
where the rotational term is similar to that of \citet{HBPWF:01}; note
that the exponent $q$ allows for sub- or super-Keplerian rotation. In 
equation 2, $M$ represents the mass of the star, $V_{\rm LPV}$ is
the stellar velocity, the angle $\theta$ is measured in the equatorial 
plane of the shell and the angle $\phi$ is the latitudinal coordinate
of the shell.    Since the mass of IK\,Tau is unknown, 1\,$M_{\odot}$ was
assumed for the model.  The stellar velocity, $V_{\rm LPV}$, was assumed
to be 35\,km\,s$^{-1}$ consistent with velocities determined from the
midpoints of the spectra of the 1612 MHz OH masers, 33.2\,km\,s$^{-1}$
\citep{BJD:89} and the H$_2$O masers 35.4\,km\,s$^{-1}$ \citep{BCJ:93}.
Similar to the models of \cite{HGYFBD:96, HGYFBD:02} maser gain length 
was approximated by considering the distance along the path in which the 
LOS velocity changes by less than three times the Doppler width of the line.
Amplification becomes negligible for velocity shifts greater than this 
limit \citep{HGYFBD:96, HGYFBD:02}.  For SiO we estimated a
Doppler line width $\Delta V_{\rm D}\approx 1$\,km\,s$^{-1}$ for a typical temperature 
of 1500\,K in the SiO maser region \citep{DGHBF:95}. 
The maser gain length was further constrained to be greater
than the minimum path length of $10^{13}$\,cm necessary for SiO
maser emission \citep{ELITZUR:82} and to be less than or equal
to the physical LOS path length through the shell.  

In practice, a random point is chosen within the region 
determined by the geometric parameters of the shell (i.e. size, orientation, 
thickness).  The LOS velocity for this point is computed from 
equation 2.  From this point, distances forwards and backwards along the 
observers line of sight are computed for which the LOS velocity changes by 
1.5 times the Doppler line width.  Thus, we have a line of points, with the 
chosen random point as the center, for which $V(r)_{\rm LOS}$ is 
within three times the Doppler line width.  If both ends of the line reside within the 
maser shell and the total length of the line is greater than the minimum
path length required for maser emission, then the center point is kept
and is color-coded according to its $V(r)_{\rm LOS}$.  Otherwise, the 
point is rejected.   In this manner, the model builds up a distribution 
of points mapping both the structure of the shell and the maser velocity
distribution.  

The model shown in Figure \ref{IKTAU_MODEL} is meant to provide a qualitative match
to the IK\,Tau SiO maser distribution.  The parameters for the model 
are presented in Table 1.  Inner distances of 16\,mas and 28\,mas were 
used for the semiminor, $A_0$, and semimajor, $B_0$, axes respectively 
with a shell thickness of 3 mas.  As mentioned previously the position 
angle for the ellipse was set to 59$^{\circ}$.  The model is not particularly
sensitive to this angle, and changes of $\lsim$5$^{\circ}$ do not change the 
results significantly.  The two parameters which affect the 
velocity structure are the exponent $q$ and the expansion velocity 
$V_{\rm exp}$.  Through an iterative process of adjusting parameters and 
qualitatively comparing the model to the data, values of $q = 1.09$ and 
$V_{\rm exp} = 3.5$\,km\,s$^{-1}$ were determined.   Coincidentally, the 
value for $q$ is equivalent to that found for R\,Aqr by \citet{HBPWF:01}.  
As in the model for R\,Aqr, the model for IK\,Tau is sensitive to 
small ($\sim$0.01) variations in the value of $q$.  Unlike the R\,Aqr case, 
the mass of IK\,Tau is not well constrained, and variations in the 
assumed value for $M$ could allow for slight variations in the value 
of $q$.  To produce a similar model to that shown in Figure \ref{IKTAU_MODEL} 
with pure Keplerian motion ($q=1$) would require a stellar mass 
$M \approx 0.1M_{\odot}$.  Similarly, the model is sensitive to the 
the expansion velocity $V_{\rm exp}$.  Changes on the order of 
$\sim$1.0\,km\,s$^{-1}$ produce significant changes in the velocity 
distributions of the masers.     

Additionally, there is a trade-off between the model expansion velocity, and the 
quenching of the masers caused by the Doppler line width limitation.  
Larger values for  $V_{\rm exp}$ tend to spread the distribution in the
middle panel of the model plot, but only if $\Delta V_{\rm D}$
is increased to allow the greater change in velocity and resulting longer
gain lengths.  The longer path lengths, in turn, create more masers in
front of and behind the star, a characteristic which is clearly not
evident in these or any other SiO maser observations of which we are
aware.  As the top-right panel of  Figure \ref{IKTAU_MODEL} shows, the 
density distribution 
of masers resulting from the path length considerations is characterized 
by tangentially amplified masers along the limb of the shell.  In order to
limit maser emission along the line of sight to the star, the 
maser shell thickness must be $\lsim$ the maser minimum path length.
Since we assumed a minimum maser path length of $10^{13}$\,cm 
($\sim$2.7\,mas at the assumed distance to IK\,Tau), the shell thickness was
set to 2.6\,mas.  Without increasing the minimum required path length, 
increasing the maser shell thickness will produce masers along the line
of sight to the star, thus filling in the distribution in the top-right panel of
Figure \ref{IKTAU_MODEL}.

The expansion velocity $V_{\rm exp} = 3.5$\,km\,s$^{-1}$ used in the 
model is consistent with observational evidence as found in proper motion 
studies of SiO masers.  \cite{BDK:97} measured the proper motions of the 
SiO masers toward R\,Aqr and found an overall infall of the masers 
(shell contraction) with a velocity of $\sim$4\,km\,s$^{-1}$ over stellar 
phases $\phi = 0.78-1.04$.  \cite{DK:03}, in an extensive study of the 
Mira variable TX\,Cam, determined the dominant mode to be expansion
of the SiO masers, with radial proper motion magnitudes of 
$\sim$5--10\,km\,s$^{-1}$ between stellar phases $\phi \approx 0.5-1.5$.  

From our single epoch of SiO maser observations, it is difficult to tell
whether the model expansion term represents a physical 
expansion or contraction of the maser shell.  This is because both 
expansion and contraction produce nearly identical distributions in the both 
the middle- and lower-right panels of Figure \ref{IKTAU_MODEL}.  
The only difference between a positive or negative expansion term is 
observable in the top-right panel of Figure \ref{IKTAU_MODEL}.   
For example, if we consider only rotation of the shell, we find that all of the 
red-shifted masers occur along the positive $X$-axis (the SE side
of the shell), and all of the blue-shifted masers occur along the 
negative $X$-axis (NW side of the shell) and the velocities are symmetric 
about the plane of the sky.  The addition of a positive expansion term 
(as was done for our model) serves to rotate the blue-shifted 
velocities out of the plane of the sky toward the front of the shell closest 
to the observer and red-shifted velocities away from the observer.  However,
the actual masers are still confined to a region around the plane of the 
sky, thus the shift in the velocity structure is most easily observed near 
the poles along the $Z$-axis where we see in the top-right panel of 
Figure \ref{IKTAU_MODEL} that some low-velocity blue-shifted masers 
(green) have been shifted toward the positive $X$-axis.   
Contraction of the shell has the opposite effect rotating the red-shifted
velocity structure out of the plane toward the front of the shell and the 
blue-shifted velocities toward the back of the shell.  It is practically
impossible to tell from our maser data whether expansion or contraction
is the case for the IK\,Tau shell.  There seems to be a slight preference 
for blue-shifted masers in general and a tendency for some blue-shifted masers 
along the positive $X$-axis in the top-left panel of Figure \ref{IKTAU_MODEL},
but this is highly dependent on the assumed stellar velocity of 35\,km\,s$^{-1}$
and the choice of position angle for the major axis of the ellipse. 

The maser shell rotational velocities produced in the model are not 
considered extraordinary.  The equatorial velocity at the distance of the mean 
semiminor axis of 16\,mas is $\sim$3.6\,km\,s$^{-1}$ with a corresponding rotational 
period of $\sim$33\,yr.  At the UD photospheric distance of 10.1~mas 
the equatorial velocity is $\sim$4.6\,km\,s$^{-1}$ with a corresponding 
rotational period of $\sim$16\,yr.  These rotational velocities are consistent 
with observational measurements for R\,Aqr 
\citep[$\sim$3--5\,km\,s$^{-1}$,][]{HBPWF:01,COTTON:04}
and slightly lower than measurements for NML\,Cyg 
\citep[$\sim$11\,km\,s$^{-1}$,][]{BM:00} and OH231.8+4.2 
\citep[$\sim$7--10\,km\,s$^{-1}$,][]{SC:02}.   Significant stellar rotation 
velocities (10s to 100s of km\,s$^{-1}$) during the 
asymptotic giant branch (AGB) phase have been invoked in theoretical 
studies of chemical mixing and the nucleosynthesis of heavy elements in 
convective envelopes \citep{LHWH:99,HLL:03,SGL:04}.   Stellar 
rotation is also factor in models describing the shaping of bipolar and 
elliptical planetary nebulae, and \cite{GLRF:99} propose that AGB stars 
above $\sim$1.3\,$M_\odot$ can spin up their extended circumstellar 
envelopes to rotational speeds on the order of a km\,s$^{-1}$.  
Stellar rotation velocities of $\sim$3--7\,km\,s$^{-1}$ are used in such 
models that result in equatorial density enhancements and eventually bipolar 
nebulae \citep{GLRF:99}.   Interestingly, the collective maser observations 
of IK Tau, which suggest rotation, axial symmetry, and equatorial density 
enhancement, may be indicating that IK Tau is on its way to becoming 
such a bipolar planetary nebula. 

\section{SUMMARY}

We have imaged the $v=1,J=1-0$, SiO maser emission toward the Mira
variable IK\,Tau.  We find that the SiO masers are distributed in an
elongated ring roughly $58\times32$\,mas ($15\times 8$\,AU at the
adopted distance of 250 pc) with the major axis of the ellipse at a
position angle of $\sim$59$^{\circ}$.  The NE--SW elongation of the 
SiO maser distribution is roughly perpendicular to the NW--SE symmetry 
axis previously determined for the H$_2$O and OH maser distributions.  This
elongation axis also provides an axis of symmetry for the LOS velocity
structure of the SiO masers.  Masers having velocities blue-shifted
with respect to the assumed center velocity of 35\,km\,s$^{-1}$ were
found to lie on the northwestern side of the shell while the
southeastern side of the shell is dominated by red-shifted SiO masers.
Under the assumption of tangentially amplified masers, this velocity
distribution implies a NW to SE rotation of the SiO shell about a NE--SW
rotation/polar axis.  Such a polar axis would be in agreement with models invoked to
explain the H$_2$O and OH maser geometries which suggest a NW--SE
equator for IK\,Tau.

The inherent transient and irregular nature of SiO maser
emission in the atmospheres of long-period variables limits the comparison 
of models to a single epoch of VLBI data.  However, we have attempted 
to approximate the IK\,Tau observations presented here assuming an ellipsoidal model 
of the SiO maser shell.  We favor a model that includes both rotation of the form 
$V \propto 1/r^{1.09}$ and constant-velocity expansion/contraction as a 
representation of the velocity field in the CSE of IK\,Tau at the distance of the 
SiO maser shell.  This model appears to reproduce
the global characteristics of the observed SiO masers in terms of structure 
and velocity distribution.  A multi-epoch VLBI monitoring program to determine 
the SiO maser proper motions could verify this scenario and would  
greatly clarify our understanding of the CSE of IK\,Tau.

\acknowledgements
The authors thank Dr. J. M. Hollis for helpful discussions during the 
preparation of this manuscript.

\clearpage 

\begin{deluxetable}{ll}
\tablecaption{ELLIPSOIDAL SHELL MODEL PARAMETERS  \label{SHELL_PARAMS}}
\tablewidth{0pt}
\tablehead{\colhead{Parameter} & \colhead{Value}}
\startdata
Shell Geometry:                                                                                        &         \\
\phantom{or}semiminor axis, $A_{0}$    \dotfill & 16\,mas   \\
\phantom{or}semimajor axis, $B_{0}$    \dotfill &  28\,mas   \\
\phantom{or}Shell thickness                          \dotfill &   2.6\,mas     \\
\phantom{or}Position angle                             \dotfill &  59$^{\circ}$   \\
\phantom{or}Inclination, $i$                          \dotfill &  0$^{\circ}$     \\
Distance                                                                         \dotfill & 250 pc   \\ 
Stellar Mass, $M$                                                      \dotfill &   1 $M_\odot$    \\
Stellar Velocity, $V_{\rm LPV}$                        \dotfill & 35.0\,km\,s$^{-1}$  \\
Minimum maser path length                              \dotfill & $1\times10^{13}$ cm  \\
Doppler line width, $\Delta V_{\rm D}$ \dotfill & 1.0\,km\,s$^{-1}$  \\
Expansion velocity, $V_{\rm exp}$                 \dotfill & 3.5\,km\,s$^{-1}$ \\
Exponent, $q$                                                                                      \dotfill & 1.09  \\
\enddata
\end{deluxetable}

\clearpage

\epsscale{0.9}

\epsscale{0.9}

\epsscale{0.9}

\epsscale{0.8}

\end{document}